\documentclass[%
 reprint,
 amsmath,amssymb,
 aps,
]{revtex4-2}

\usepackage{booktabs}
\usepackage{braket}
\usepackage{physics}
\usepackage{graphicx}% Include figure files
\usepackage{dcolumn}% Align table columns on decimal point
\usepackage{bm}% bold math
\usepackage{hyperref}
\usepackage{float}
\hypersetup{
  colorlinks=true,
  linkcolor=blue,
  citecolor=blue,
  urlcolor=blue
}
\begin{document}

\preprint{APS/123-QED}

\title{PINNs for Solving Two-Flavor Neutrino Oscillations in Vacuum and Matter Environments for Atmospheric and Reactor Neutrinos; A detailed analysis}

\author{T. Srinivasan}
\email{srinivasan.t2022@vitstudent.ac.in}
\affiliation{Department of Physics, School of Advanced Sciences, Vellore Institute of Technology, Chennai, India}

\author{Kalyani Desikan}
\email{kalyanidesikan@vit.ac.in}
\affiliation{Department of Mathematics, School of Advanced Sciences, Vellore Institute of Technology, Chennai, India}

\date{\today}

\begin{abstract}
Neutrino oscillations provide crucial insights into fundamental particle physics, with two-flavor approximations offering a robust framework for analyzing reactor and atmospheric phenomena. Currently, standard numerical integrators represent the gold standard for solving the associated evolution equations, achieving machine precision at negligible computational cost. However, Physics-Informed Neural Networks (PINNs) have recently emerged as a powerful surrogate model for handling high-dimensional systems, complex geometries, and inverse parameter estimation. In this paper, we establish a foundational proof-of-concept by applying PINNs to solve the coupled ordinary differential equations governing neutrino evolution in vacuum and constant-density matter profiles. We also discuss the theoretical framework encompassing vacuum mixing and the matter-enhanced neutrino oscillations and rigorously benchmark the PINNs against standard NuFast and RK4 (Runge-Kutta 4) solutions. By carefully optimizing our network architecture, we demonstrate that PINNs can achieve high-precision agreement with experimental standards for reactor (low-energy) and atmospheric (high-energy) neutrinos, which we use as our case study. This paper evaluates the robustness of the PINNs approach through a comprehensive error analysis, establishing a baseline validation for its application to more challenging problems in neutrino physics, such as variable density profiles.
\end{abstract}

\maketitle

\section{Introduction}

\label{intro}
Neutrinos, elusive particles with tiny masses, exhibit flavor oscillations due to quantum mixing between mass eigenstates, a phenomenon confirmed by experiments in laboratories such as Super-Kamiokande (atmospheric) \cite{Fukuda1998} and the Sudbury Neutrino Observatory (SNO) (solar) \cite{Ahmad2002}. Although the three-flavor framework provides the most complete description, the two-flavor approximation remains a robust and essential tool for understanding dominant oscillation channels \cite{Giunti2007}, such as those observed in solar, atmospheric and reactor neutrino experiments ~\cite{Eguchi2003,Abe2011,An2012}.

In the two-flavor picture, one typically considers the \(\nu_e \leftrightarrow \nu_\mu\) (or \(\nu_e \leftrightarrow \nu_\tau\)) system driven by the solar mass splitting \(\Delta m_{21}^2 \sim 7.5\times 10^{-5}~\text{eV}^2\) and mixing angle \(\theta_{12} \simeq 34^\circ\), or the \(\nu_\mu \leftrightarrow \nu_\tau\) system driven by the atmospheric mass splitting \(\Delta m_{32}^2 \sim 2.5\times 10^{-3}~\text{eV}^2\) and mixing angle \(\theta_{23} \simeq 45^\circ\) ~\cite{PDG2020,Esteban2020,deSalas2021}.

Reactor antineutrinos are produced as a byproduct of the nuclear fission process in a nuclear reactor. In a typical pressurized water reactor (PWR) or boiling water reactor (BWR), energy is generated primarily through thermal neutron-induced fission of four primary isotopes: $^{235}$U, $^{238}$U, $^{239}$Pu, and $^{241}$Pu \cite{Mueller2011}. When these heavy nuclei undergo fission, they split into two lighter, neutron-rich fission fragments. To reach the valley of beta stability, these unstable fragments undergo a series of beta decays ($\beta^-$), emitting an electron and an electron antineutrino in the process.
\begin{equation}
_{Z}^{A}X \rightarrow \: _{Z+1}^{A}Y + e^- + \bar{\nu}_e
\end{equation}

 The energy of these antineutrinos typically ranges from $0$ to about $10 - 12 \text{ MeV}$\cite{Giunti2007}, peaks at approximately $2 \text{ MeV}$. Reactor neutrino experiments have provided precise measurements of the mixing angle $\theta_{13}$ and confirmed the three-flavor oscillation paradigm~\cite{An2012,RENO2012,An2016}. These results are consistent with solar and atmospheric neutrino observations~\cite{Fukuda1998,Ahmad2002}. Reactor neutrinos offer a precise and controlled probe of neutrino mixing parameters in the vacuum regime. And they play a crucial role in completing the global picture of neutrino oscillations~\cite{GonzalezGarcia2012}.

Atmospheric neutrinos are generated when high-energy cosmic rays collide with air nuclei \cite{Gaisser2002,Kajita2010}, in the atmosphere of the earth producing pions ($\pi\pm$) and kaons (K$\pm$) that decay into muons ($\mu\pm$) and neutrinos ($\nu\pm$). Its neutrino oscillations are driven predominantly by the atmospheric mass-squared splitting and near-maximal mixing. It covers an immense range of baselines \(L\) and energies \(E\): neutrinos can be observed from directly overhead (\(L\sim 15\) km) to trajectories through the Earth's diameter (\(L\sim 1.2\times 10^4\) km), with energies from sub-GeV to TeV scales \cite{Akhmedov2000}. Unlike reactor neutrinos, atmospheric neutrinos can traverse Earth, experiencing matter effects in the mantle and core compared to the former. Determining precise oscillation probabilities through these dense environments is crucial to determine the ordering of the neutrino mass \cite{Abe2017,Acero2019}.

Traditional numerical methods, such as Runge--Kutta (RK4) or finite difference schemes \cite{Press2007}, have long been the golden standard for solving neutrino evolution equations. Although effective for constant densities and simple profiles, mesh-based approaches face significant challenges when applied to realistic, highly variable density profiles like the Preliminary Reference Earth Model (PREM) \cite{Dziewonski1981}, often requiring adaptive step-sizes and complex interpolation \cite{Kopp2006} to maintain stability near density discontinuities (e.g., the core-mantle boundary). Furthermore, they make the inverse problems such as inferring oscillation parameters from data computationally expensive \cite{Morales2026}.

In contrast, \emergencystretch=2em Physics-Informed Neural Networks (PINNs) \cite{Raissi2019} represent a paradigm shift in computational physics. This Machine Learning based approach \cite{Carleo2019} offers a promising method for solving the evolution equations of neutrino oscillations by incorporating physical constraints directly into neural network training, achieving high accuracy by eliminating computational mesh requirements. Unlike traditional purely data-driven approaches, PINNs integrate fundamental physical laws encoded as partial differential equations (PDEs) directly into the neural network loss function, enabling a very close approximation to achieve accurate solution even in data-sparse regimes. This paradigm shift represents a significant departure from conventional numerical methods such as finite element methods (FEM) and finite difference methods (FDM)\cite{Karniadakis2021,Cuomo2022}, which rely on explicit mesh generation and suffer from computational overhead in high-dimensional problems. The PINNs framework achieves physical consistency by incorporating governing PDEs into the loss function as soft constraints, thereby regularizing the learned solution and preventing overfitting to noisy measurements. This approach has proven highly effective in fields ranging from fluid dynamics \cite{Raissi2020,Jin2021} to quantum mechanics \cite{Karniadakis2021,Lagaris1998}. The primary advantage of PINNs lies in their flexibility, particularly their ability to evaluate governing equations at randomly sampled continuous points, entirely bypassing the computational bottleneck of generating discrete grids for variable density profiles. 

Furthermore, PINNs seamlessly facilitate data assimilation by embedding empirical measurements - such as observed survival probabilities from neutrino detectors - directly into the network's loss function alongside the physical constraints. This architecture naturally allows the simultaneous solution of forward and inverse problems \cite{Cuomo2022,Lu2021}. Although PINNs have been successfully applied to fluid dynamics and classical mechanics, their application to quantum unitary evolution in high-energy physics remains underexplored. This paper presents a new approach to PINNs in the domain of neutrino phenomenology. Unlike previous general-purpose solvers, we particularly design a complex-valued network architecture specifically tailored to preserve the unitarity of the neutrino evolution operator. In addition, since standard deep neural networks only output real numbers, the PINNs cannot directly output a single complex number. So, we split the complex Schr\"odinger equation into its real and imaginary components here, which would be discussed further in detail.

The remainder of this paper is organized as follows. In Section~\ref{sec:formalism}, we review the theoretical formalism of two-flavor neutrino oscillations, detailing the quantum mechanical propagation dynamics in both vacuum and constant matter environments. Section~\ref{sec:methodology} presents the methodology of the proposed Physics-Informed Neural Network (PINNs), including the network architecture, the treatment of the complex Schr\"odinger equation, and the construction of the physics-informed loss function. In Sections~\ref{sec:results1} and \ref{sec:results2}, we present our numerical results and hence discuss the detailed performance of the PINNs framework in capturing the survival and appearance probabilities for reactor and atmospheric neutrino configurations and also compared with standard solvers such as NuFast\cite{Denton2020} and Runge-Kutta 4 (RK-4) methods in detail. Finally, Section~\ref{sec:conclusion} concludes the paper and outlines potential directions for future research.

\section{The Formalism of Neutrino Oscillations }\label{sec:formalism}
%%\label{}
In this section, we outline the two-flavor neutrino oscillations in vacuum and constant matter environments by detailing the mathematical formalisms for its propagation.

\subsection{Two-flavor oscillations in vacuum}
The neutrino flavor eigenstates \(\ket{\nu_\alpha}\) (with \(\alpha = e,\mu\)) are related to the mass eigenstates \(\ket{\nu_i}\) (\(i=1,2\)) via a unitary mixing matrix \cite{Xing2011,Maki1962},
\begin{equation}
\begin{pmatrix}
\ket{\nu_e} \\
\ket{\nu_\mu}
\end{pmatrix}
=
\begin{pmatrix}
\cos\theta & \sin\theta \\
-\sin\theta & \cos\theta
\end{pmatrix}
\begin{pmatrix}
\ket{\nu_1} \\
\ket{\nu_2}
\end{pmatrix},
\end{equation}
where \(\theta\) is the vacuum mixing angle and \(\Delta m^2 = m_2^2 - m_1^2\) is the mass-squared difference between the mass eigenstates \(\nu_2\) and \(\nu_1\).
The time evolution of the mass eigenstates is governed by a Schr\"odinger-like equation,
\begin{equation}
i \frac{d}{dt} \ket{\nu_i(t)} = E_i \ket{\nu_i(t)},
\end{equation}
with \(E_i \approx p + m_i^2/(2E)\), where \(p\) and \(E\) are the momentum and energy of neutrinos with \(p \approx E\). In flavor space, the evolution can be written as \cite{Xing2011}
\begin{equation}
    i \frac{d}{dt} \begin{pmatrix} \ket{\nu_e(t)} \\ \ket{\nu_\mu(t)} \end{pmatrix} = H_F \begin{pmatrix} \ket{\nu_e(t)} \\ \ket{\nu_\mu(t)} \end{pmatrix},
    \label{eq:state_evolution}
\end{equation}
where \(x \approx t\) for ultra-relativistic neutrinos and the effective Hamiltonian on the flavor basis is given by the equation;
\begin{equation}
H_{\text{F}} = \frac{\Delta m^2}{4E}
\begin{pmatrix}
-\cos 2\theta & \sin 2\theta \\
\sin 2\theta & \cos 2\theta
\end{pmatrix}.
\end{equation} 
Any arbitrary state \(\ket{\psi(t)}\) that starts as a pure electron neutrino state at \(t=0\) can be expanded in terms of the pure flavor eigenstates \(\ket{\nu_e}\) and \(\ket{\nu_\mu}\), where: \begin{equation}
   \ket{\psi (t)} =\psi _{ee}(t) \ket{\nu _{e}} +\psi _{e\mu }(t)\ket{\nu _{\mu }}, \end{equation} 
with the probability amplitudes of \(\ket{\nu_{\alpha}} \to \ket{\nu_{\beta}}\) transitions after a time \(t\) is given by
\(\psi_{\alpha\beta}(t) = \langle \nu_{\beta} | \nu_{\alpha}(t) \rangle\). Now, by substituting the effective Hamiltonian matrix H$_F$ in the time evolution equation for the two flavor neutrino oscillation in terms of the mixing angle, gives \cite{Giunti2007}; 
\begin{equation}
i \frac{d}{dx}
\begin{pmatrix}
\psi_{ee} \\
\psi_{e\mu}
\end{pmatrix}
=
\frac{1}{4E}
\begin{pmatrix}
-\Delta m^{2}\cos 2\theta & \Delta m^{2}\sin 2\theta \\
\Delta m^{2}\sin 2\theta & \Delta m^{2}\cos 2\theta
\end{pmatrix}
\begin{pmatrix}
\psi_{ee} \\
\psi_{e\mu}
\end{pmatrix},
\end{equation} 
with the initial conditions given by
 \begin{equation}
\Psi_e(0)
=
\begin{pmatrix}
\psi_{ee}(0) \\
\psi_{e\mu}(0)
\end{pmatrix}
=
\begin{pmatrix}
1 \\
0
\end{pmatrix}.
\label{eq:initial}
\end{equation}
The appearance and survival probabilities at baseline \(L\) are found using the relations;
\begin{align}
P_{e\mu}^{\text{vac}}(L) = |\psi_{e\mu}(x)|^2, \quad \\ 
P_{ee}^{\text{vac}}(L)=|\psi_{ee}(x)|^2=1-P_{e\mu}^{\text{vac}}(L) 
\end{align}
The analytical solution for the evolution equation is then given by the transition probabilities; 
\begin{align}
P_{ee}^{\text{vac}}(L) &= 1 - \sin^2 2\theta \, \sin^2\left(\frac{\Delta m^2 L}{4E}\right), \label{eq:app}\\
P_{e\mu}^{\text{vac}}(L) &= \sin^2 2\theta \, \sin^2\left(\frac{\Delta m^2 L}{4E}\right).
\label{eq:surv}
\end{align}
These probabilities depend on the ratio \(L/E\), leading to the characteristic oscillatory behavior used in experimental analyses.

\subsection{Two-flavor oscillations in matter}
When neutrinos propagate through matter, the coherent forward scattering of \(\nu_e\) on electrons via charged-current interactions introduces an effective potential \cite{Wolfenstein1978}
\begin{equation}
V_{\text{CC}} = \sqrt{2} G_F n_e,
\end{equation}
where \(G_F = 1.166\times 10^{-5}~\text{GeV}^{-2}\) is the Fermi constant and \(n_e\) is the electron number density of the matter. For electrically neutral matter, neutral-current contributions are flavor-universal and can be dropped in the two-flavor approximation.

The effective Hamiltonian in matter becomes
\begin{equation}
H_{\text{mat}} =
\frac{\Delta m^2}{4E}
\begin{pmatrix}
-\cos 2\theta & \sin 2\theta \\
\sin 2\theta & \cos 2\theta
\end{pmatrix}
+
\begin{pmatrix}
V_{\text{CC}} & 0 \\
0 & 0
\end{pmatrix}.
\end{equation}

For constant matter density, it is convenient to introduce the dimensionless parameter
\begin{equation}
Acc = \frac{2\sqrt{2} G_F n_e E}{\Delta m^2},
\end{equation}
where \(n_e \simeq 3.65\times 10^{27}~\text{m}^{-3}\). The effective mixing angle in matter \(\theta_M\) and effective mass-squared difference \(\Delta m_M^2\) become,
\begin{align}
\sin^2 2\theta_M &= \frac{\sin^2 2\theta}{(\cos 2\theta - Acc)^2 + \sin^2 2\theta}, \\
\Delta m_M^2 &= \Delta m^2 \sqrt{(\cos 2\theta - Acc)^2 + \sin^2 2\theta}.
\end{align}
In our study, we consider the constant matter density of the earth core ($\rho \approx$ 13 $ g/cm^3$) for neutrinos to travel. The evolution equation of neutrinos to oscillate between two flavors in terms of the mixing angle in matter is given as \cite{Giunti2007}
\begin{multline}
i \frac{d}{dx}
\begin{pmatrix}
\psi_{ee} \\
\psi_{e\mu}
\end{pmatrix}
=\\
\frac{1}{4E}
\begin{pmatrix}
-\Delta m^{2}_{\mathrm{M}} \cos 2\theta_{\mathrm{M}} &
\Delta m^{2}_{\mathrm{M}} \sin 2\theta_{\mathrm{M}} \\
\Delta m^{2}_{\mathrm{M}} \sin 2\theta_{\mathrm{M}} &
\Delta m^{2}_{\mathrm{M}} \cos 2\theta_{\mathrm{M}}
\end{pmatrix}
\begin{pmatrix}
\psi_{ee} \\
\psi_{e\mu}
\end{pmatrix}.
\end{multline}

After performing the orthogonal transformation and diagonalising the effective Hamiltonian matrix H$_{mat}$ the evolution equation becomes;
$\Psi_e = U_M \Phi_e$,\\ with \(\Psi_e = \begin{pmatrix} \psi_{ee} \\ \psi_{e\mu} \end{pmatrix}, \quad \Phi_e = \begin{pmatrix} \phi_{e1} \\ \phi_{e2} \end{pmatrix}\) \\\text{and} \( U_M = \begin{pmatrix}
\cos \theta_M & \sin \theta_M \\
-\sin \theta_M & \cos \theta_M
\end{pmatrix}\).\\
If the matter-density is constant then the $d\theta_M/dx$ = 0 and from the initial conditions~(\ref{eq:initial}) we have
\begin{equation}
\begin{pmatrix}
\phi_{e1}(0) \\
\phi_{e2}(0)
\end{pmatrix}
=
\begin{pmatrix}
\cos \theta_M & -\sin \theta_M \\
\sin \theta_M & \cos \theta_M
\end{pmatrix}
\begin{pmatrix}
1 \\
0
\end{pmatrix}
=
\begin{pmatrix}
\cos \theta_M \\
\sin \theta_M
\end{pmatrix}.
\end{equation}

The oscillation probabilities in constant matter density are then
\begin{align}
P_{ee}^{\text{mat}}(L) &= 1 - \sin^2 2\theta_M \, \sin^2\left( \frac{\Delta m_M^2 L}{4E} \right) \\
P_{e\mu}^{\text{mat}}(L) &= \sin^2 2\theta_M \, \sin^2\left( \frac{\Delta m_M^2 L}{4E} \right).
\end{align}

For realistic propagation through the Earth or Sun, the density varies with position. In that case, \(H(x)\) becomes explicitly position-dependent via \(n_e(x)\), and the evolution equation must be solved with a space-dependent Hamiltonian, often using the PREM profile for Earth \cite{Dziewonski1981} or an exponential model for the Sun \cite{Bahcall1989} respectively.

\section{Methodology of Physics-Informed Neural Networks}\label{sec:methodology}
This section details our computational methodology. We first describe the neural network architecture designed to output the real and imaginary components of the wavefunction. Then formulate the composite physics-informed loss function that rigorously enforces the Schr\"odinger equation-based evolution and initial boundary conditions, and finally discuss the specific implementations of the Hamiltonian for both vacuum and matter propagations.

\subsection{Network architecture}

We approximate the complex-valued neutrino wavefunction \(\psi(x)\) using a fully connected feed-forward neural network. The primary input is the evolution parameter (baseline \(x\) or time \(t\)), and the outputs are the real and imaginary parts of the flavor amplitudes. For the two-flavor case, we write
\begin{equation}
\psi(x) =
\begin{pmatrix}
\psi_e(x) \\
\psi_\mu(x)
\end{pmatrix}
=
\begin{pmatrix}
\psi_{e}^{\text{Re}}(x) + i \psi_{e}^{\text{Im}}(x) \\
\psi_{\mu}^{\text{Re}}(x) + i \psi_{\mu}^{\text{Im}}(x)
\end{pmatrix},
\end{equation}
and let the network output the four real components
\begin{equation}
\mathbf{u}(x) =
\left(
\psi_{e}^{\text{Re}},
\psi_{e}^{\text{Im}},
\psi_{\mu}^{\text{Re}},
\psi_{\mu}^{\text{Im}}
\right).
\end{equation}

The architecture used in our analyses consists of 6 hidden layers with 128–256 neurons per layer, employing hyperbolic tangent (\(\tanh\)) activation function \cite{Wang2020}. Training is carried out using gradient-descent based optimizers such as Adam \cite{Kingma2014} and/or L-BFGS \cite{Rafati2018}. Adam is often employed in the early training stages for robust exploration of the parameter space, while L-BFGS can refine the solution and improve convergence near minima \cite{Rathore2024,Mustajab2024}. The \(\tanh\) activation function is preferred over ReLU \cite{Sitzmann2020,Raissi2019} for physics-informed problems due to its smoothness, allowing for the computation of higher-order derivatives required by the physics loss \cite{Jagtap2020}. To overcome the inherent "spectral bias" of standard deep neural networks, we incorporated a Fourier feature projection at the network's input layer. We selected this Fourier biasing approach over alternative methods, because it mathematically preconditions the input coordinate space while fully preserving the robustness and gradient stability of standard activations (like tanh) during complex L-BFGS optimization. Weights are initialized using Glorot (Xavier) normal initialization to maintain stable gradients at the start of training.

\subsection{Physics-informed loss function}

The PINNs is trained by minimizing a composite loss function $\mathcal{L}_{\text{total}}$ \cite{Raissi2019}:
\begin{equation}
\mathcal{L}_{\text{total}} = \lambda_\text{ODE}\mathcal{L}_{\text{ODE}} + \lambda_\text{BC}\mathcal{L}_{\text{BC}}.
\end{equation}

where \(\mathcal{L}_{\text{ODE}}\) enforces the evolution equation and normalization by means of probability conservation. \(\mathcal{L}_{\text{BC}}\) enforces the initial boundary conditions, such as the initial flavor state. And its, corresponding weights are given as \(\lambda_\text{ODE} \ \text{and} \ \lambda_\text{BC}\) respectively.

The evolution equation in vacuum or matter can be written generically as
\begin{equation}
i \frac{d}{dx} \psi(x) - H(x)\psi(x) = 0.
\end{equation}

\begin{figure*}[htbp]
    \centering
    \includegraphics[width=0.9\linewidth]{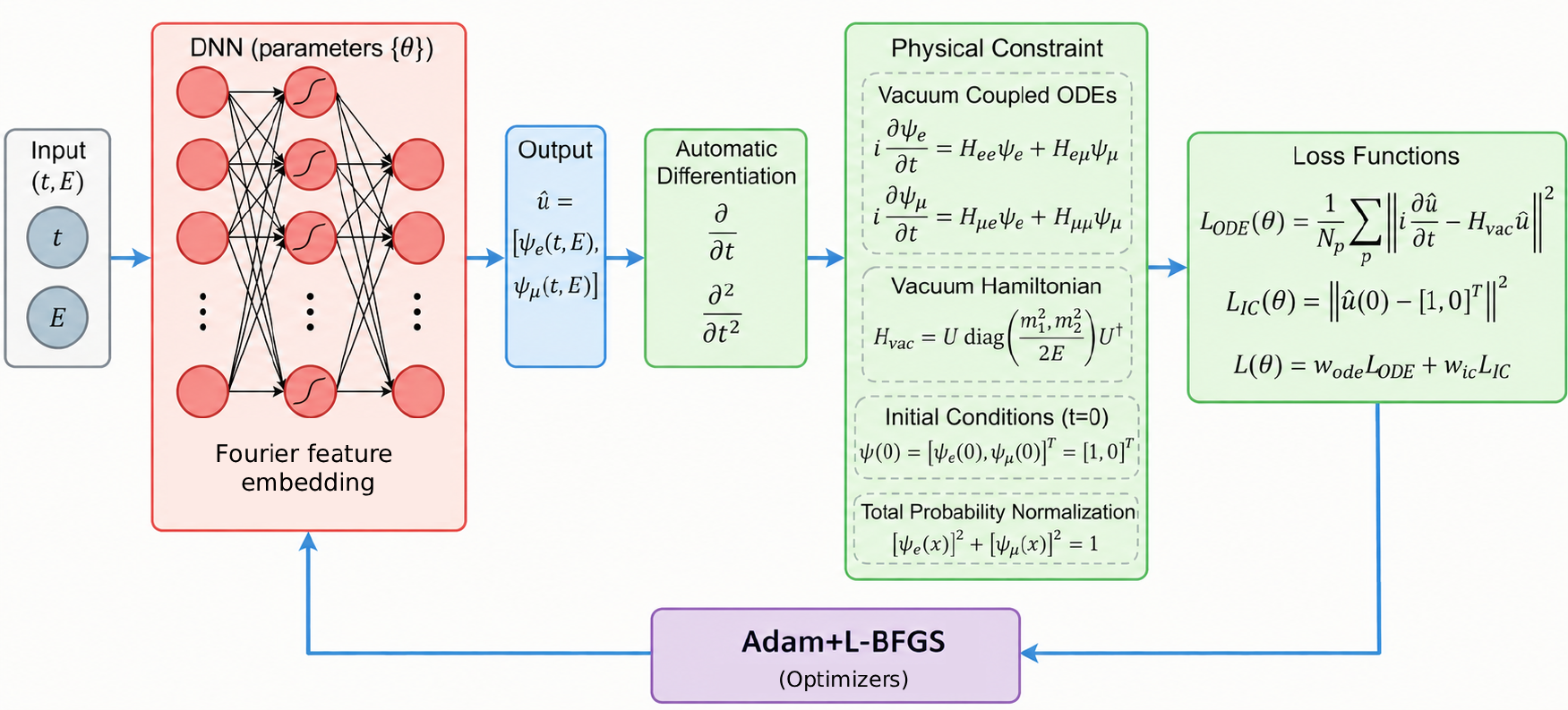}
    \caption{PINNs architecture for neutrino propagation in vacuum.}
    \label{fig:my_label}
\end{figure*}
\begin{figure*}[htbp]
    \centering
    \includegraphics[width=0.9\linewidth]{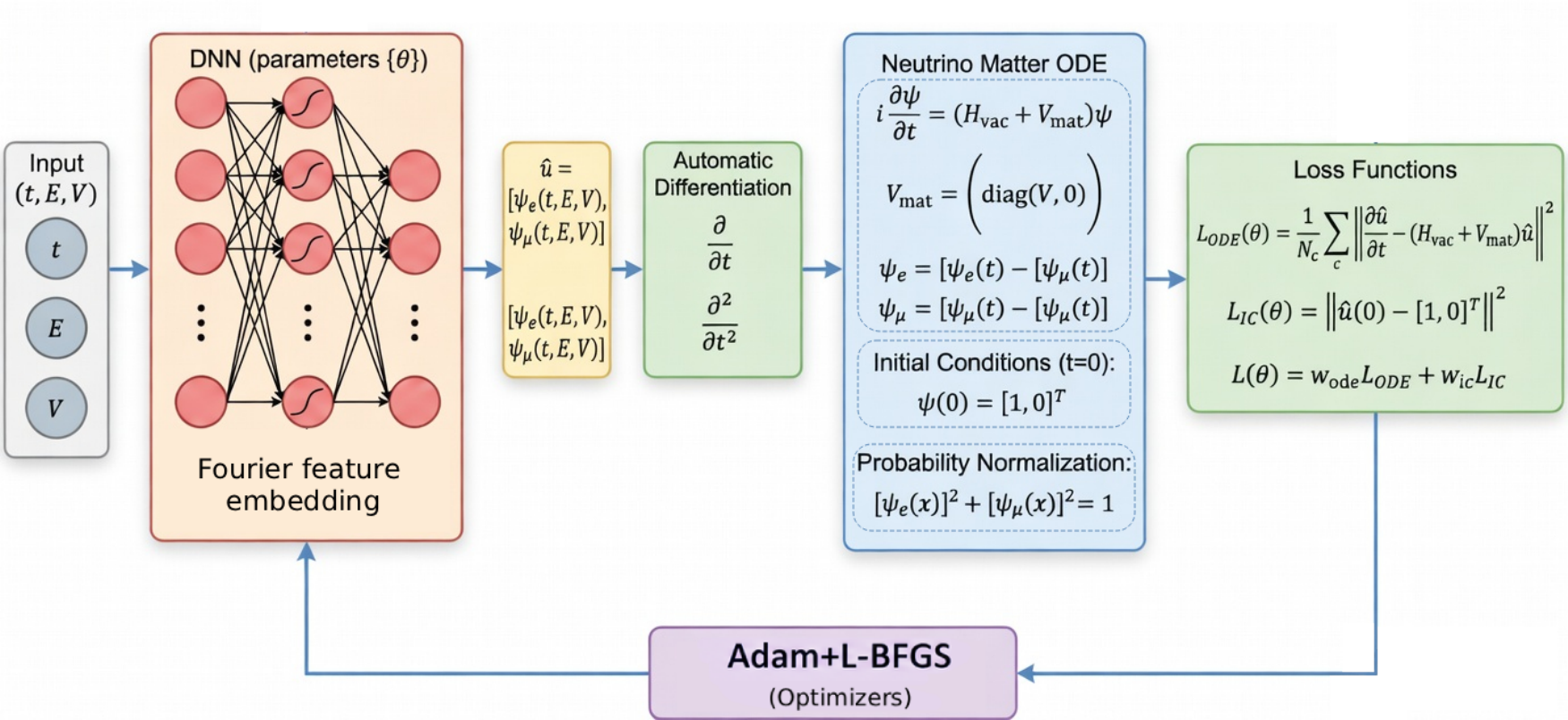}
    \caption{PINNs architecture for neutrino propagation in matter.}
    \label{fig:my_label}
\end{figure*}

In practice, we split the real and imaginary parts. For example, in vacuum the flavor Hamiltonian is
\begin{equation}
H_{\text{vac}} = \frac{\Delta m^2}{4E}
\begin{pmatrix}
-\cos 2\theta & \sin 2\theta \\
\sin 2\theta & \cos 2\theta
\end{pmatrix},
\end{equation}
and the corresponding real–imaginary system can be written as four coupled real equations for \(\psi_{e}^{\text{Re}},
\psi_{e}^{\text{Im}},\psi_{\mu}^{\text{Re}}\) and \(\psi_{\mu}^{\text{Im}}\). The residuals of these equations at collocation points \(\{x_i\}\) define \(\mathcal{L}_{\text{ODE}}\).
\begin{equation}
\mathcal{L}_{\text{ODE}} = \frac{1}{N_f} \sum_{i=1}^{N_f} \left\| i \frac{d{\psi}}{dx}(x_i) - \hat H {\psi}(x_i) \right\|^2,
\end{equation}
where \(N_f\) is the number of collocation sample points taken. Automatic differentiation (AD) \cite{Paszke2019}, is used here, to compute spatial derivatives such as \(d\psi_{e}^{\text{Re}}/dx\) exactly (to machine precision)\cite{Baydin2018}, avoiding discretization error and maintaining a mesh-free formulation. Also, to maintain the perfect oscillating nature of the equation, the total probability is normalized throughout the evolution:
\begin{equation}
|\psi_e(x)|^2 + |\psi_\mu(x)|^2 = 1.
\end{equation}
Normalization can be enforced either as a soft constraint in the loss or monitored as a diagnostic of training quality.

The boundary-condition loss \(\mathcal{L}_{\text{BC}}\) enforces that at the 
production point \(x=0\), the neutrino state is a pure flavor, for example
\begin{equation}
\psi_e(0) = 1, \qquad \psi_\mu(0) = 0,
\end{equation}
and is therefore defined as
\begin{equation}
\mathcal{L}_{\mathrm{BC}}=\left\|\Psi_{\mathrm{NN}}(0)-\begin{pmatrix}1 \\0
\end{pmatrix}
\right\|^2,
\end{equation}
where \(\Psi_{NN}(0)\) is the initial condition given by equation~(\ref{eq:initial}).
\subsection{Implementation details}

The main difference between vacuum and matter propagation in PINNs is the Hamiltonian that is used in the residual. In vacuum, \(H(x) = H_{\text{vac}}\) does not depend on matter density for fixed \(\Delta m^2, E, \theta\). For constant matter density, we use
\begin{equation}
H_{\text{mat}} = H_{\text{vac}} +
\begin{pmatrix}
V_{\text{CC}} & 0 \\
0 & 0
\end{pmatrix},
\end{equation}
with \(V_{\text{CC}}\) constant. The vacuum PINNs architecture for neutrinos oscillating in 2-flavors is given in Fig. 1.

For spatially varying profiles, such as PREM for Earth \cite{Dziewonski1981} or an exponential solar profile, we use
\begin{equation}
H(x) = H_{\text{vac}} + V_{\text{CC}}(x),
\end{equation}
with the electron density \(n_e(x)\) given by the chosen model. The distinguishing feature for the matter case lies in the physics-informed loss function which forces the network to discover the non-linear amplitude enhancement and frequency shifts associated with the MSW effect. Figure 2 shows the PINNs architecture for the constant matter propagation of neutrinos oscillating in 2-flavors. 

The network thus learns to approximate oscillations in both vacuum and matter once the analytic expressions for \(\theta_M\) or \(\Delta m_M^2\) are given; and after few trials and errors, these emerge from the governing differential equation.\\

\section{Reactor Neutrino Configuration - Results and discussion}\label{sec:results1}

In this section, we present the numerical solutions for two-flavor neutrino oscillations obtained using the Physics-Informed Neural Network (PINNs) framework. We evaluate the network's performance for the \textbf{Reactor Neutrino} regime (low energy, small mass splitting). And then, we compare the PINNs predictions against the standard solutions in vacuum and constant matter environments with 
meticulous error analysis. The reactor neutrino oscillations were simulated using the characteristic parameters of the \(\nu_e \to \nu_\mu\) channel: neutrino energy \(E = 0.015~\text{GeV}\) (\(15~\text{MeV}\)), solar mass-squared difference \(\Delta m^2_{21} = 7.41 \times 10^{-5}~\text{eV}^2\), and mixing angle \(\theta_{12} = 33.4^\circ\).
\subsection{Vacuum Propagation}
 The PINNs accurately reproduces the non-maximal oscillation amplitude characteristic of the specified mixing angle, correctly resolving the survival probability floor at $P_1 \approx 0.15$ and the appearance probability ceiling at $P_2 \approx 0.85$ as seen in Fig. 3. The model demonstrates phase alignment across multiple oscillation cycles exhibiting no artificial damping, or frequency distortion throughout the cycle. The RMSE values for Survival Probability ($P_{\nu_e \rightarrow \nu_e}$) $7.3456 \times 10^{-2}$ and Appearance Probability($P_{\nu_e \rightarrow \nu_\mu}$) $8.1410 \times 10^{-2}$ confirm the consistent global convergence within an $\mathcal{O}(10^{-2})$ error regime across both channels.

\begin{figure*}[htbp]
    \centering
    \includegraphics[width=0.9\linewidth]{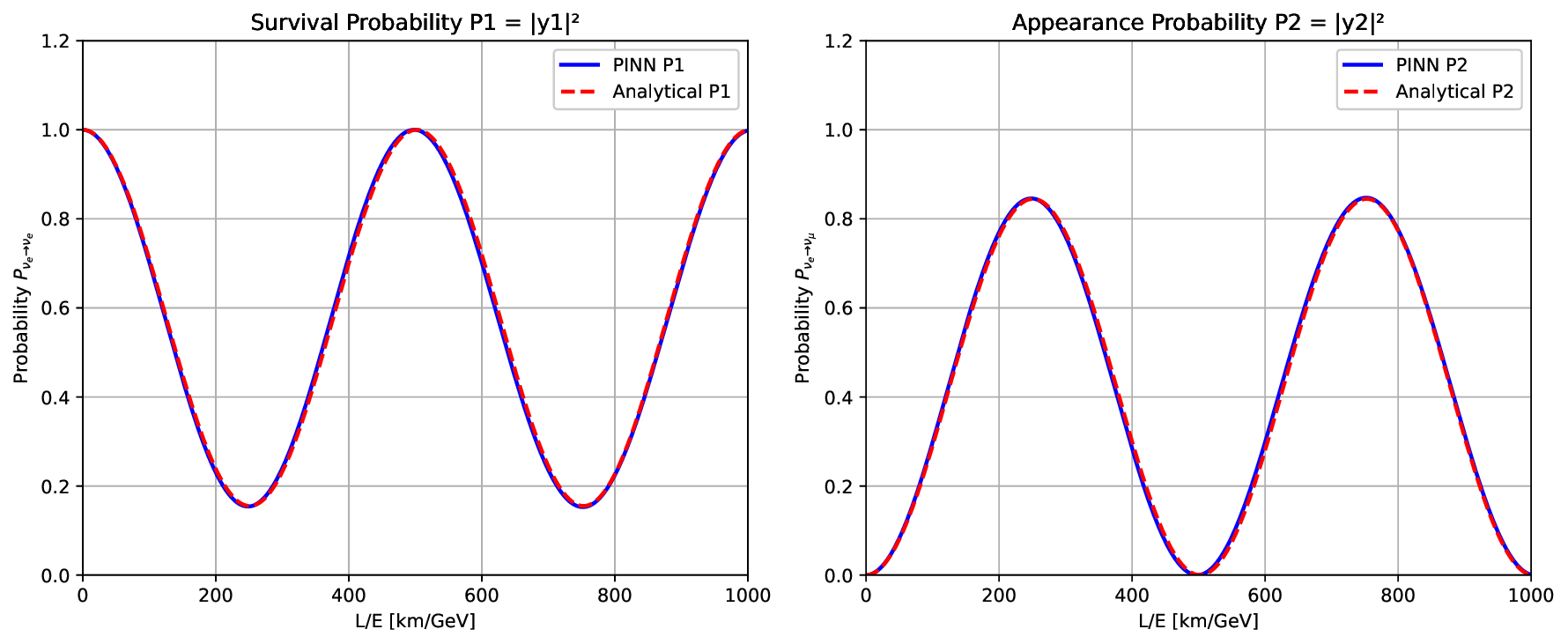}
    \caption{Survival and Appearance Probabilities of Reactor neutrino in vacuum}
    \label{fig:placeholder}
\end{figure*}
\begin{figure*}[htbp]
    \centering
    \includegraphics[width=0.9\linewidth]{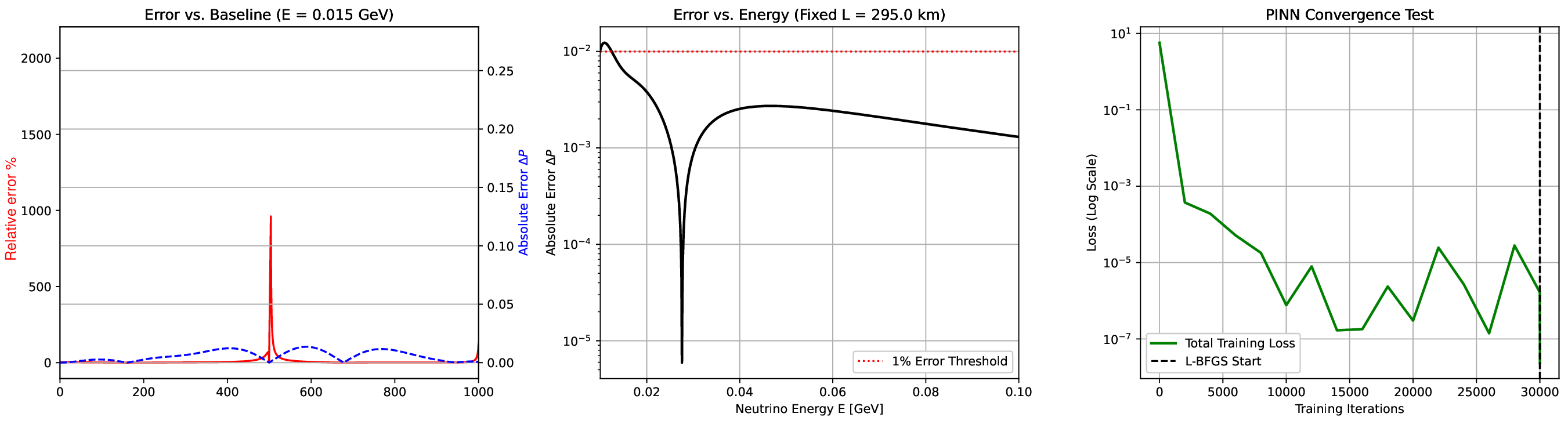}
    \caption{PINNs performance estimations from the Baseline error, Energy error analysis and the Convergence test for reactor neutrinos propagating in vacuum}
    \label{fig:placeholder}
\end{figure*}
\subsubsection{Error analysis and convergence test}
This section deals with the Baseline and Energy error analysis of reactor neutrinos in vacuum, followed by the convergence test details showing its effectiveness in problem solving (refer Fig. 4).\\
\textbf{Baseline Error Dynamics:} ($E = 0.015\text{ GeV}$)\\
Global Precision: Across the $1000\text{ km}$ spatial domain, the continuous absolute error $\Delta P$ (blue dashed curve) remains highly bounded, maintaining sub-percent to $1.5\%$ accuracy ($\Delta P < 0.015$) for the vast majority of the trajectory.\\ Localized Divergence: A singular, severe deviation manifests as a sharp red spike precisely at the oscillation node near $L \approx 500\text{ km}$, peaking at approximately $\Delta P \approx 0.125$ ($12.5\%$). Because this spike is strictly localized and the surrounding baseline error remains minimal, it represents a fractional phase misalignment exactly at the mathematical turning point rather than a widespread failure of the network's predictive capability.\\
\textbf{Energy Error Dynamics:} (Fixed $L = 295.0\text{ km}$)\\
The error climbs above the $1\%$ dotted threshold ($10^{-2}$) for neutrino energies below $0.015\text{ GeV}$. Since decreasing $E$ increases the $L/E$ ratio and vastly accelerates the oscillation frequency, the neural network exhibits a classic "spectral bias"—struggling to preserve phase accuracy for high-frequency waves. As $E$ climbs past $0.03\text{ GeV}$ (lowering the oscillation frequency), the absolute error monotonically decays down toward $10^{-3}$, demonstrating that the PINNs excels at broader, lower-frequency extrapolations.\\
\textbf{Optimization and Convergence Tests:}\\
Adam Phase ($0$–$30,000$ iterations): The network navigates the loss landscape efficiently over the full $30,000$-step Adam training phase, traversing six orders of magnitude from above $10^{0}$ down to the $10^{-7}$ regime. The stochastic jitter observed throughout this phase (including the spikes near $15,000$, $20,000$, and just before $30,000$) is characteristic of the first-order Adam optimizer actively escaping local minima. \\
L-BFGS Phase ($30,000$ iterations onwards): The activation of the second-order L-BFGS optimizer exactly at iteration $30,000$ drives the sudden, sharp vertical plunge visible at the extreme right edge of the plot. L-BFGS rapidly exploits the local curvature to smooth the remaining gradients, clamping the loss down to a deeply converged minimum. 

\subsubsection{Benchmarking PINNs Against RK4 and NuFast}
The predictive accuracy of the Physics-Informed Neural Network (PINNs) was benchmarked against two established computational standards: a 4th-order Runge-Kutta (RK4) integrator and the high-performance C++ framework NuFast.\\
In the vacuum regime, all three independent computational approaches - NuFast C++ (solid light blue band), RK4 Python numerical integration (solid blue line), and the PINNs surrogate (dashed red line) - exhibit exact visual superposition as shown in Fig.5. The network displays negligible cumulative phase drift or amplitude degradation over multiple complete oscillation periods ($\lambda_L \approx 500\text{ km}$), confirming that the physics-informed loss accurately preserves the vacuum Hamiltonian dynamics.\\

\begin{figure*}[htbp]
    \centering
    \includegraphics[width=0.9\linewidth]{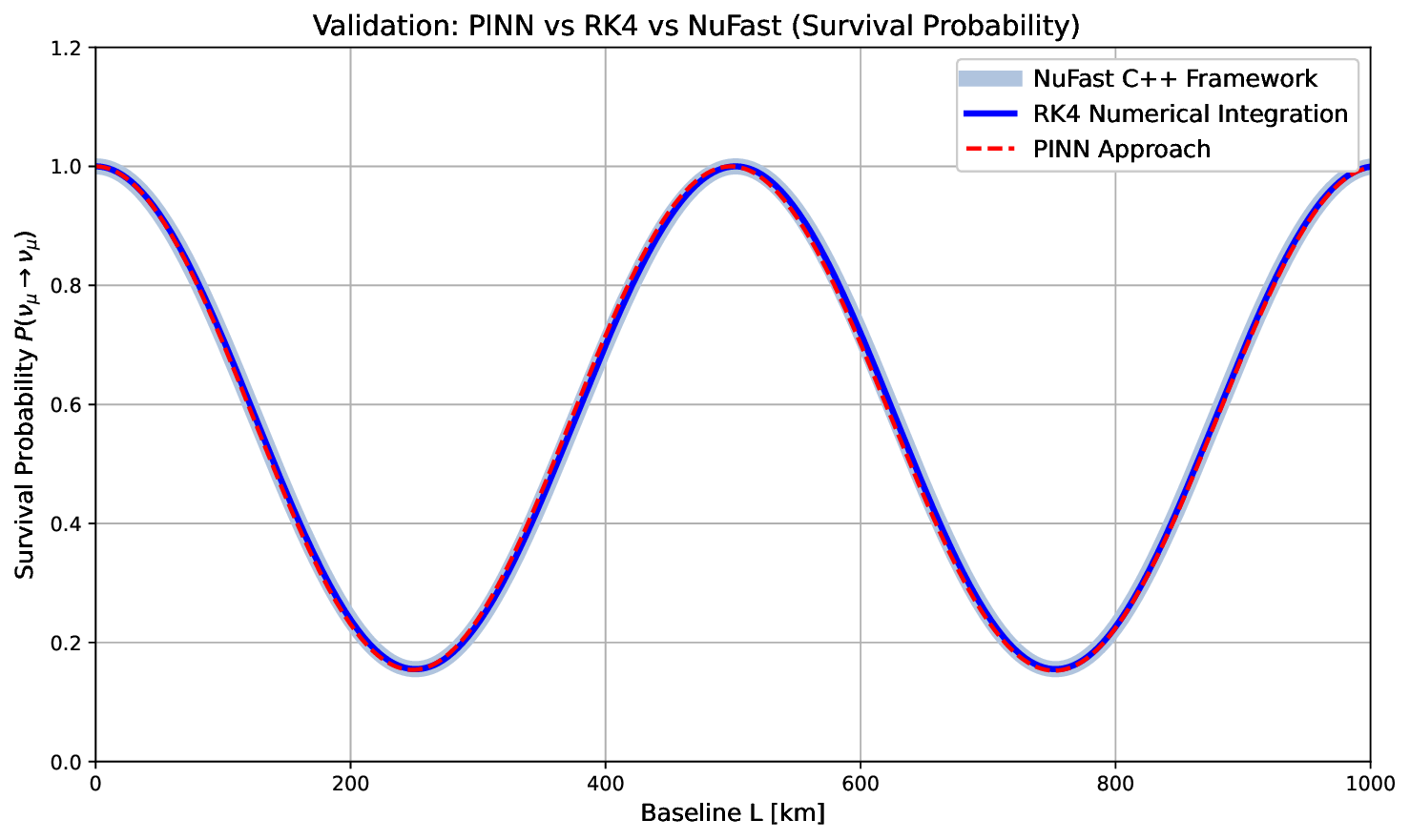}
    \caption{Benchmarking PINNs for the reactor neutrinos oscillating in vacuum versus RK4 and NuFast algorithms under similar oscillating conditions}
    \label{fig:placeholder}
\end{figure*}

\textbf{Average relative errors from Benchmark algorithms:}\\
Table 1 presents a quantitative comparison of the survival probabilities generated by the PINNs against numerical solvers (RK4 and NuFast). The Mean Absolute Error (MAE) demonstrates exceptional agreement, with the average divergence.
\begin{table}[H]
\centering
\caption{Mean Absolute error of PINNs compared with RK4 and NuFast for the reactor neutrinos in vacuum.}
\label{tab:mae_comparison}
\small
\begin{tabular}{lc}
\toprule
\textbf{Comparison} & \textbf{MAE} \\
\midrule
PINNs vs RK4   & $5.8681\times10^{-3}$ \\
PINNs vs NuFast & $5.8683\times10^{-3}$ \\
\bottomrule
\end{tabular}
\end{table}

\subsection{Constant Matter Propagation}
To investigate matter effects, we introduce a constant electron density potential \(V\). The inclusion of the matter potential alters the effective mass-squared difference, fundamentally shifting the oscillation frequency. As seen from Fig. 6, the PINNs perfectly adapts to this modified wavelength ($\lambda_{L/E} \approx 540$ km/GeV, with distinct extrema near $270$ and $810$ km/GeV) and maintains strict phase alignment with the analytical ground truth across the entire domain. The calculated RMSE values; Survival Probability ($P_{\nu_e \rightarrow \nu_e}$) RMSE: $5.6471 \times 10^{-2}$ and Appearance Probability ($P_{\nu_e \rightarrow \nu_\mu}$) RMSE: $5.1226 \times 10^{-2}$ indicate that the graph appears nearly flawless, with errors of order $\mathcal{O}(10^{-2})$.
\begin{figure*}[htbp]
    \centering
    \includegraphics[width=0.9\linewidth]{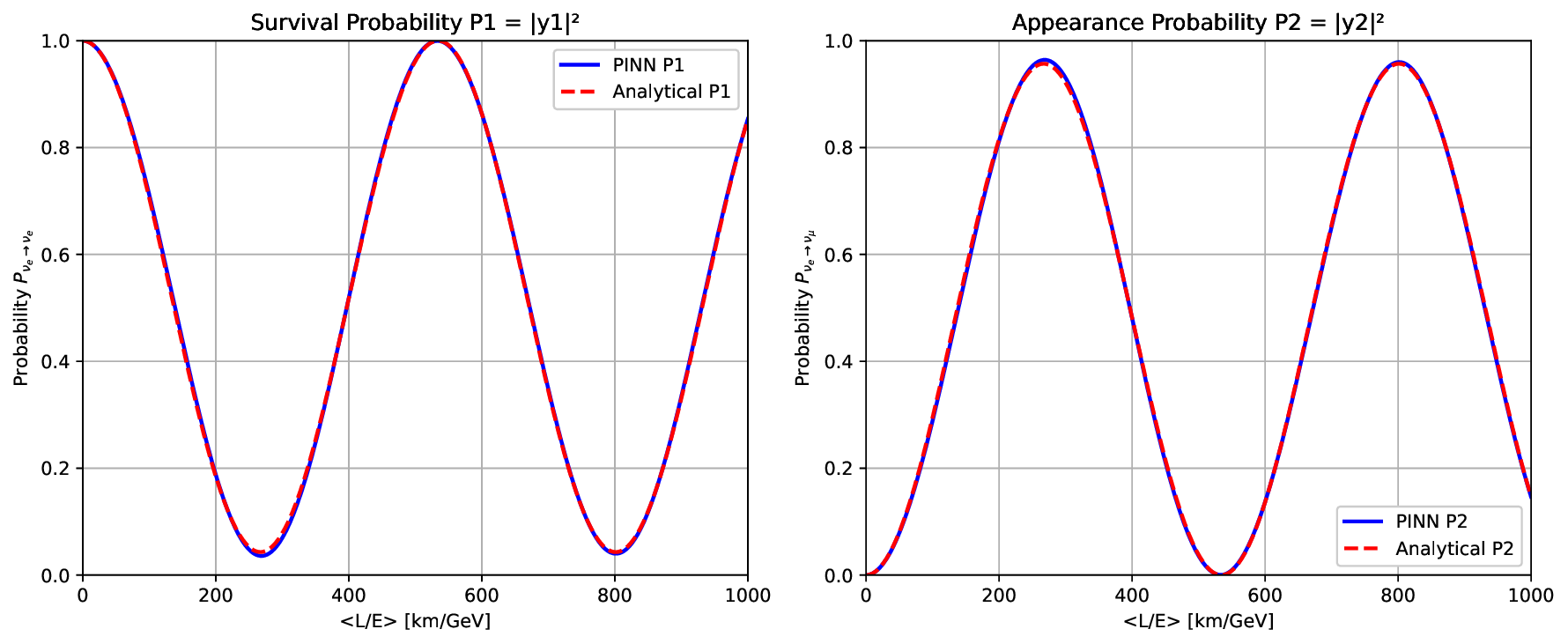}
    \caption{Survival and Appearance Probabilities of Reactor neutrino in constant matter}
    \label{fig:placeholder}
\end{figure*}
\subsubsection{Error analysis and convergence test}
This section deals with the Baseline and Energy error analysis of reactor neutrinos in constant-matter, followed by the convergence test details showing its effectiveness in problem solving (refer Fig. 7).\\
\textbf{Baseline Error Dynamics:} ($E = 0.015\text{ GeV}$)\\
Global Boundary: Across the evaluated $1000\text{ km}$ spatial domain, the absolute error $\Delta P$ (blue dashed curve) remains relatively stable, oscillating with a bounded maximum amplitude of roughly $\Delta P \approx 0.035$ ($3.5\%$).\\
Localized Divergence: A singular, severe deviation manifests as a sharp red spike precisely at $L \approx 535\text{ km}$, peaking at $\Delta P \approx 0.10$ ($10\%$). The spike highlights a fractional phase misalignment rather than a failure of the network's predictive capability at its local node density.\\
\textbf{Energy Error Dynamics:} (Fixed $L = 295.0\text{ km}$)\\
The network struggles across the evaluated energy spectrum ($0.01$ to $0.10\text{ GeV}$) when extrapolated, with the absolute error residing well above the $1\%$ threshold ($10^{-2}$ red dotted line). Since the PINNs was explicitly trained only on the domain $L/E \in [0, 1000]\text{ km/GeV}$, these high errors ($20\%$ to $40\%$) are the direct mathematical result of the neural network being forced to extrapolate massively outside its training boundaries.\\
\textbf{Optimization and Convergence Tests:}\\
Adam Phase ($0$–$30,000$ iterations): The first-order Adam optimizer efficiently traverses the loss landscape, dropping the total training loss from above $10^{1}$ down to a plateau near $10^{-7}$. The stochastic jitter observed (particularly near $15,000$ and $28,000$ iterations) is standard for this optimizer as it escapes local minima.\\
L-BFGS Phase ($30,000$ iterations onwards): The transition to the second-order L-BFGS optimizer precisely at iteration $30,000$ triggers an immediate, vertical descent, clamping the loss down to a deeply converged global minimum ($< 10^{-8}$). 
\begin{figure*}[htbp]
    \centering
    \includegraphics[width=0.9\linewidth]{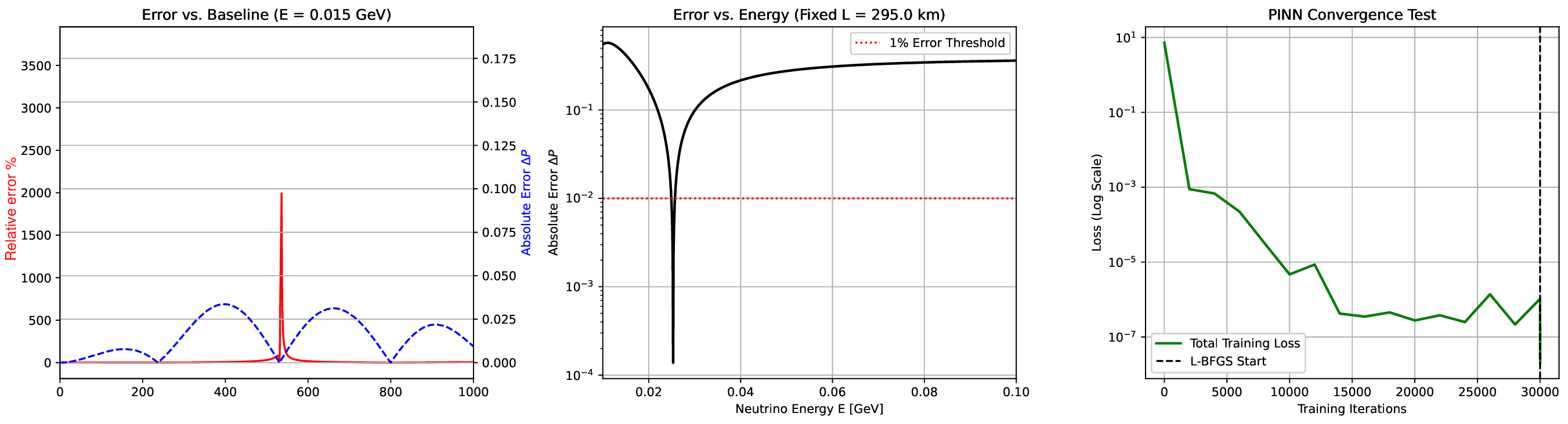}
    \caption{PINNs performance estimations from the Baseline error, Energy error analysis and the Convergence test for reactor neutrinos propagating in constant matter}
    \label{fig:placeholder}
\end{figure*}
\begin{figure*}[htbp]
    \centering
    \includegraphics[width=0.9\linewidth]{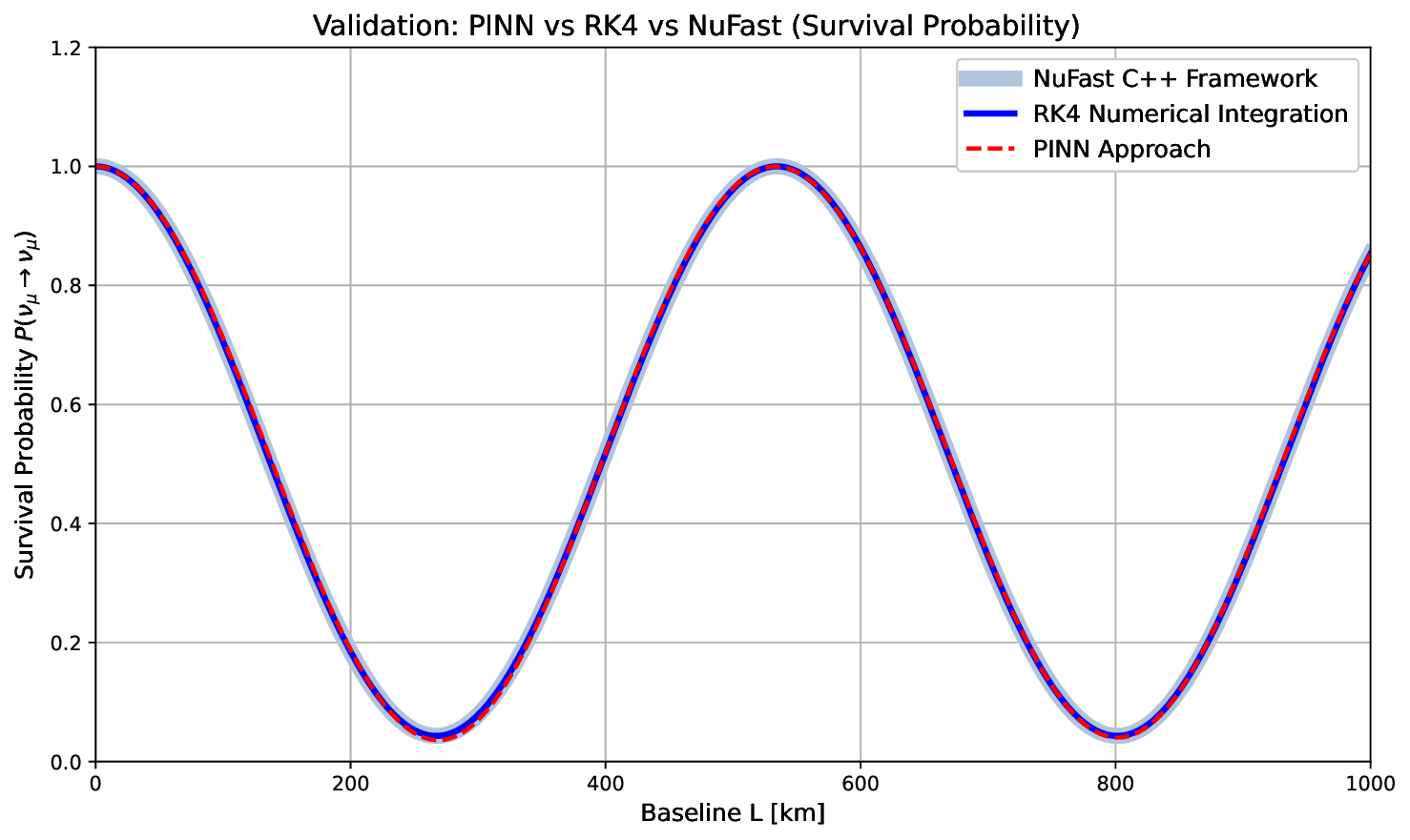}
    \caption{Benchmarking PINNs for the reactor neutrinos oscillating in constant matter versus RK4 and NuFast algorithms under similar oscillating conditions}
    \label{fig:placeholder}
\end{figure*}
\subsubsection{Benchmarking PINNs Against RK4 and NuFast}
Under the constant matter profile, the PINNs trajectory remains tightly bounded within the envelope defined by NuFast and RK4 across the entire domain, demonstrating that the network handles the additional matter potential terms without introducing boundary errors or unphysical oscillations (see Fig. 8).\\

\textbf{Average relative errors from Benchmark algorithms:}\\
As summarized in Table 2, a quantitative comparison of the Mean Absolute Error (MAE) demonstrates strong concordance between our neural network approach and standard computational baselines. The PINNs achieves an average error of exactly $1.3722 \times 10^{-2}$ when compared to both RK4 and NuFast, confirming that the network reliably captures the neutrino oscillation physics.
\begin{table}[H]
\centering
\caption{Mean Absolute error of PINNs compared with RK4 and NuFast for the reactor neutrinos in constant matter.}
\label{tab:mae_comparison}
\small
\begin{tabular}{lc}
\toprule
\textbf{Comparison} & \textbf{MAE} \\
\midrule
PINNs vs RK4   & $1.3722\times10^{-2}$ \\
PINNs vs NuFast & $1.3722\times10^{-2}$ \\
\bottomrule
\end{tabular}
\end{table}

\section{Atmospheric Neutrino Configuration - Results and discussion}\label{sec:results2}
The atmospheric neutrino simulations were conducted using the \(\nu_\mu \to \nu_\tau\) parameters: higher energy \(E = 0.2~\text{GeV}\), larger atmospheric mass splitting \(\Delta m^2_{32} = 2.4 \times 10^{-3}~\text{eV}^2\), and maximal mixing angle \(\theta_{23} = 45^\circ\). In this section, we evaluate the network's performance for the \textbf{Atmospheric Neutrino} regime (higher energy, larger mass splitting). And then, we compare the PINNs predictions against the standard solutions in vacuum and constant matter environments with robust error analysis. 

\subsection{Vacuum Propagation}
In the atmospheric vacuum case, the oscillation is governed by near-maximal mixing (\(\sin^2 2\theta \approx 1\)) as shown in  Fig. 9.
\begin{figure*}[htbp]
    \centering
    \includegraphics[width=0.9\linewidth]{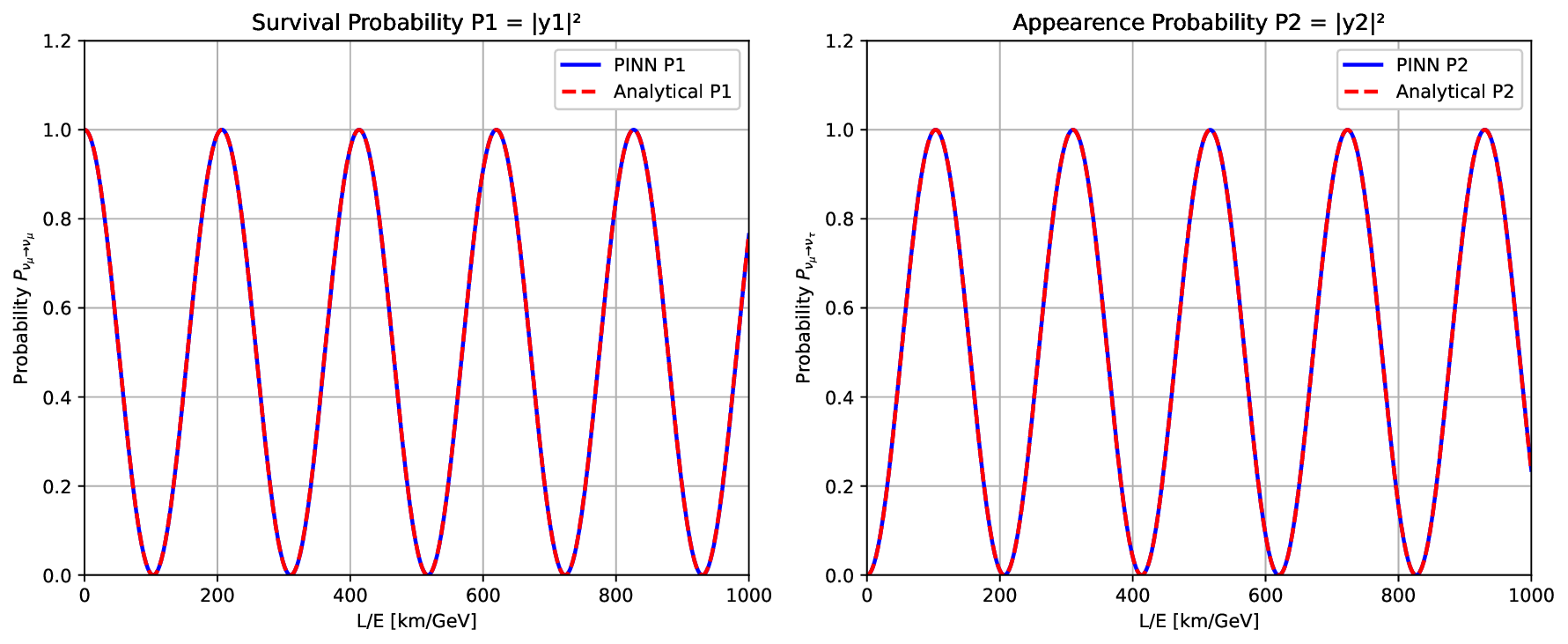}
    \caption{Survival and Appearance Probabilities of Atmospheric neutrino in vacuum}
    \label{fig:placeholder}
\end{figure*}
The PINNs (solid blue line) accurately captures the characteristic sinusoidal frequency and unit amplitude of the analytical solution (dashed red line), which confirms the network's ability to accurately capture the maximal mixing dynamics characteristic of the atmospheric sector. However, the model achieved a Root Mean Square Error (RMSE) of $1.5931 \times 10^{-3}$ for the survival probability ($P_{\nu_\mu \rightarrow \nu_\mu}$) and $1.2685 \times 10^{-3}$for the appearance probability ($P_{\nu_\mu \rightarrow \nu_\tau}$). These $\mathcal{O}(10^{-3})$ error bounds represent a highly converged global minimum. This magnitude of precision confirms that the PINNs successfully bypasses spectral bias, strictly enforcing both the underlying Hamiltonian kinematics and the continuous unitarity constraint ($P_1 (survival) + P_2(appearence) = 1$) without suffering from localized boundary spikes at the oscillation nodes, suggesting favourable network training dynamics.\\

\subsubsection{Error analysis and convergence test}
This section deals with the Baseline and Energy error analysis of atmospheric neutrinos in vacuum, followed by the convergence test details showing its effectiveness in problem solving (refer Fig. 10).\\
\textbf{Baseline Error Dynamics ($E = 0.2$ GeV)}\\
Across the $1000$ km spatial domain, the absolute error $\Delta P$ (blue dashed curve) exhibits highly stable, bounded oscillations. The maximum deviation peaks at roughly $3.5 \times 10^{-3}$ ($0.35\%$) near the domain boundary ($L \approx 980$ km). Unlike standard network failures at high-gradient turning points, the error remains exceptionally low precisely at the nodal crossings. This confirms the PINNs successfully resolves the maximal mixing kinematics of atmospheric vacuum oscillations without localized phase drift.\\
\textbf{Energy Error Dynamics ($L = 295.0\text{ km}$)}\\
The model demonstrates exceptional broadband generalization. Across the entire evaluated energy spectrum ($0.1$ to $2.0$ GeV), the absolute error remains strictly below the $1\%$ threshold ($10^{-2}$ dotted line). At lower energies ($E < 0.5$ GeV), which correspond to highly oscillatory regimes with large $L/E$ ratios, the error experiences minor spectral bias but is successfully capped at $\approx 3 \times 10^{-3}$. For $E > 1.0$ GeV, the error smoothly stabilizes near $10^{-3}$.\\ 
\textbf{Optimization and Convergence Tests:}\\
Adam Phase ($0$–$30,000$ iterations): The first-order optimizer systematically minimizes the total training loss across seven orders of magnitude, dropping from roughly $10^1$ to a plateau near $10^{-7}$. Standard stochastic jitter is visible (notably near $24,000$ iterations) as the algorithm navigates local minima.\\
L-BFGS Phase ($30,000$ iterations onwards): The transition to the second-order L-BFGS optimizer at the $30,000$ iteration mark induces an immediate vertical plunge in the loss function. This final stage clamps the loss to a deeply converged minimum, which directly accounts for the absence of localized baseline spikes and the strict adherence to sub-$1\%$ error bounds across the energy spectrum.
\begin{figure*}[htbp]
    \centering
    \includegraphics[width=0.9\linewidth]{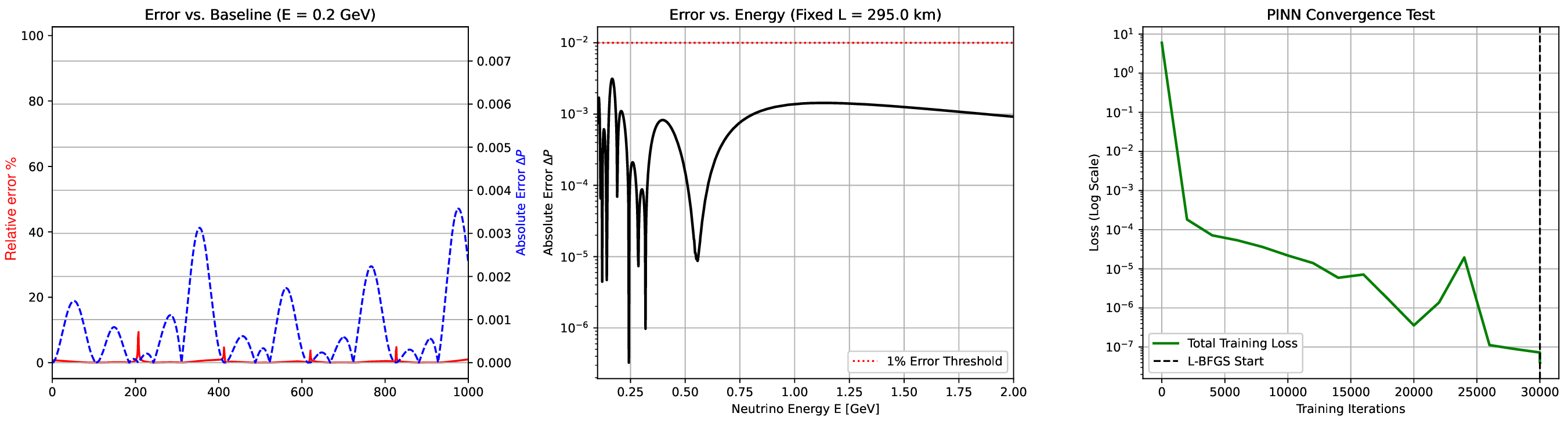}
    \caption{PINNs performance estimations from the Baseline error, Energy error analysis and the Convergence test for atmospheric neutrinos propagating in vacuum}
    \label{fig:placeholder}
\end{figure*}
\subsubsection{Benchmarking PINNs Against RK4 and NuFast}
The comparison evaluates the continuous survival probability ($P_{\nu_\mu \to \nu_\mu}$) of atmospheric neutrinos across a spatial baseline $L \in [0, 1000]\text{ km}$ for vacuum propagation. The PINNs (dashed red line) tracks both the NuFast C++ solver (solid light blue band) and the RK4 integrator (solid blue line), capturing the maximal mixing dynamics with complete amplitude swings between $1.0$ and $0.0$. The network preserves the fundamental oscillation wavelength ($\lambda_L \approx 210\text{ km}$ at $E = 0.2\text{ GeV}$) over nearly five complete cycles as seen in Fig. 11. It exhibits zero observable phase lag, frequency distortion, or artificial numerical damping across the entire $1000\text{ km}$ domain. The PINNs accurately resolves the sharp turning points at the minima ($L \approx 105, 315, 525, 735, 945\text{ km}$) and returns to unity at the nodal peaks without overshooting physical bounds.\\

\begin{figure*}[htbp]
    \centering
    \includegraphics[width=0.9\linewidth]{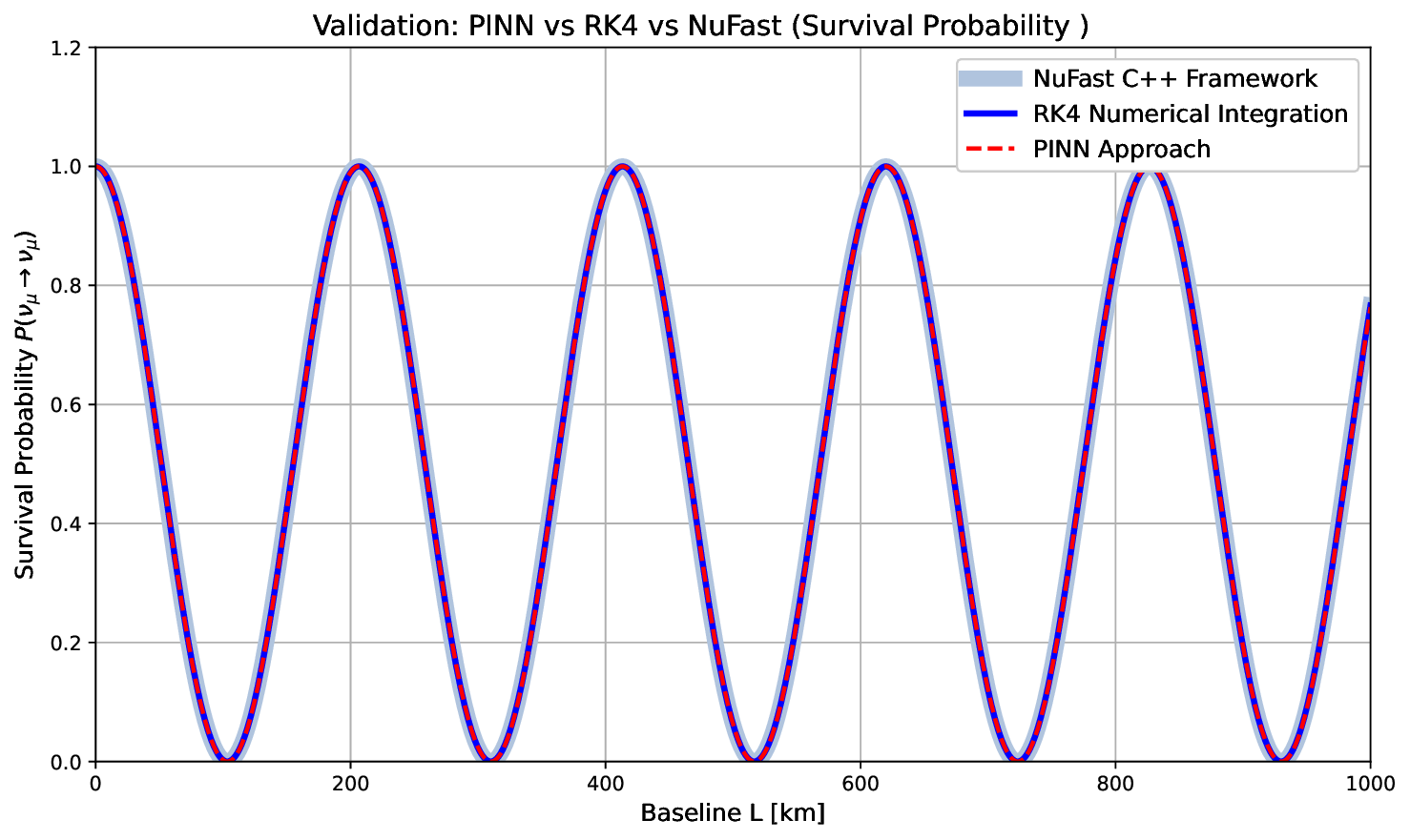}
    \caption{Benchmarking PINNs for the atmospheric neutrinos oscillating in vacuum versus RK4 and NuFast algorithms under similar oscillating conditions}
    \label{fig:placeholder}
\end{figure*}
\textbf{Average relative errors from Benchmark algorithms:}\\
Table 3 details the Mean Absolute Error (MAE) between the PINN's predicted survival probabilities and classical numerical solvers. The PINNs demonstrates exceptional precision, achieving an MAE of $1.0221 \times 10^{-3}$ against RK4 and $1.0228 \times 10^{-3}$ against NuFast.

\begin{table}[H]
\centering
\caption{Mean Absolute error of PINNs compared with RK4 and NuFast for the atmospheric neutrinos in vacuum.}
\label{tab:mae_comparison}
\small
\begin{tabular}{lc}
\toprule
\textbf{Comparison} & \textbf{MAE} \\
\midrule
PINNs vs RK4   & $1.0221\times10^{-3}$ \\
PINNs vs NuFast & $1.0228\times10^{-3}$ \\
\bottomrule
\end{tabular}
\end{table}

\subsection{Constant Matter Propagation}
\begin{figure*}[htbp]
    \centering
    \includegraphics[width=0.9\linewidth]{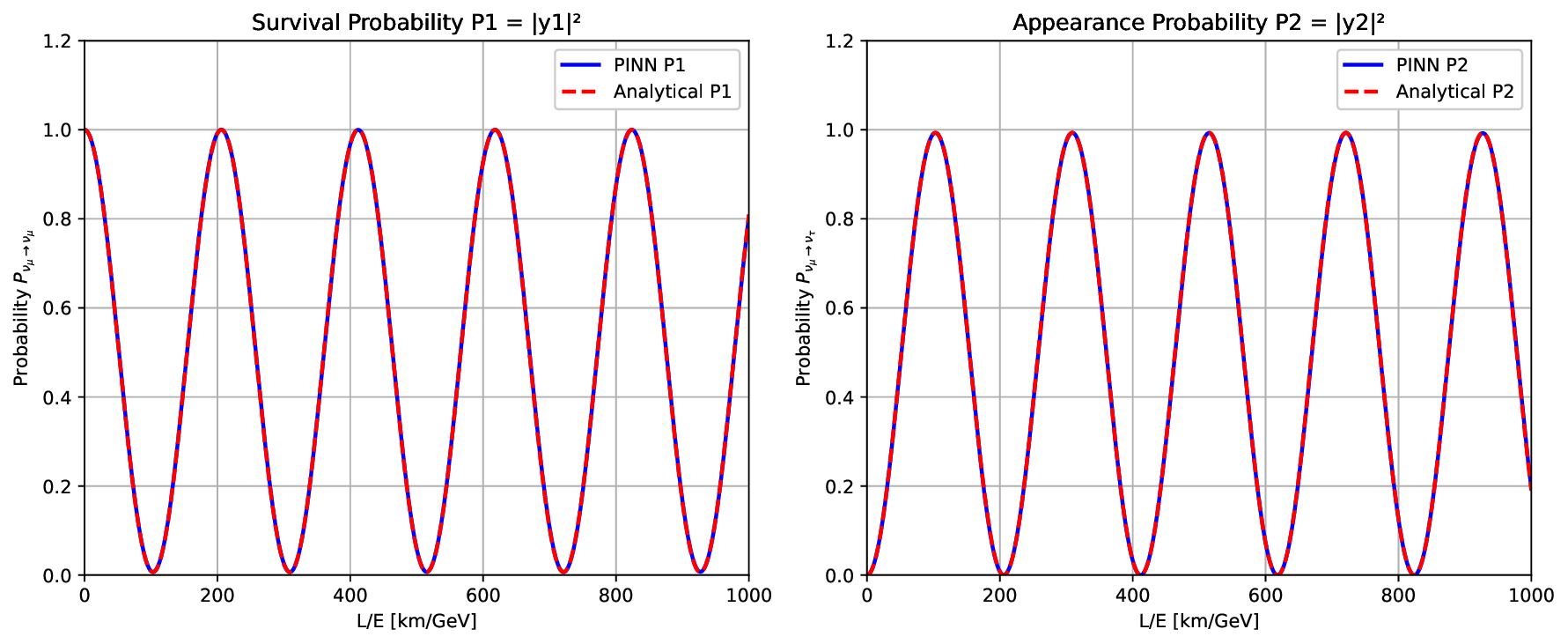}
    \caption{Survival and Appearance Probabilities of Atmospheric neutrino in constant matter}
    \label{fig:placeholder}
\end{figure*}
\begin{figure*}[htbp]
    \centering
    \includegraphics[width=0.9\linewidth]{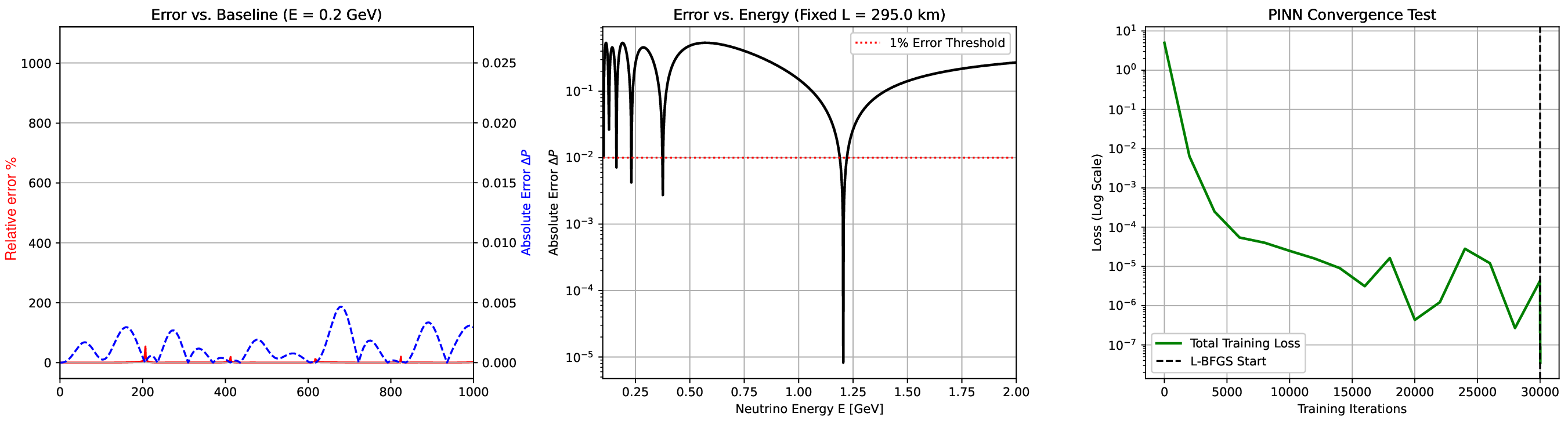}
    \caption{PINNs performance estimations from the Baseline error, Energy error analysis and the Convergence test for atmospheric neutrinos propagating in constant matter}
    \label{fig:placeholder}
\end{figure*}
In the case of matter, at \(E=0.2~\text{GeV}\), the matter potential \(V\) is relatively small compared to the vacuum term \(\Delta m^2/2E\). Consequently, the matter effect is less pronounced compared to reactor case, in the resonance region. The PINNs effectively reproduces this behavior, generating a solution that deviates only slightly from the vacuum case, consistent with theoretical expectations for sub-GeV atmospheric neutrinos in standard earth densities with the root mean squared error of around $9.2909 \times 10^{-3}$ for survival and around $7.3764 \times 10^{-3}$ for appearance probability, respectively as shown in Fig. 12. Operating well within an $\mathcal{O}(10^{-3})$ error regime confirms that the PINN's optimization successfully reached a highly converged global minimum. This sub-1\% global deviation indicates that the network efficiently encodes the matter-modified Hamiltonian dynamics.

\begin{figure*}[htbp]
    \centering
    \includegraphics[width=0.9\linewidth]{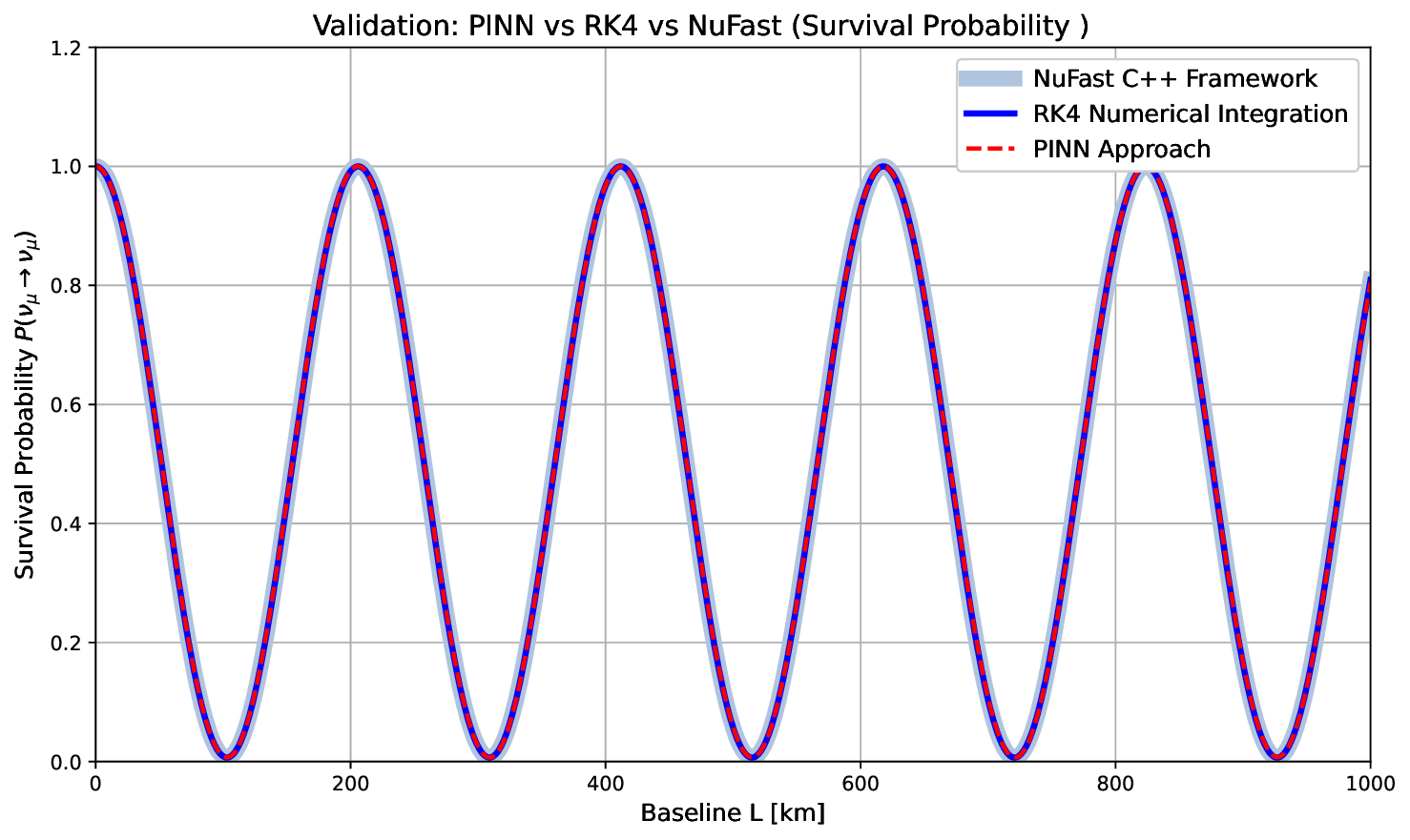}
    \caption{Benchmarking PINNs for the atmospheric neutrinos oscillating in constant matter versus RK4 and NuFast algorithms under similar oscillating conditions}
    \label{fig:placeholder}
\end{figure*}
The results confirm that the PINNs architecture can be adapted to handle the complex-valued coupled ODE system without numerical instability, provided the hyperparameters, particularly learning rate and iteration count, are tuned to mitigate spectral bias.\\

\subsubsection{Error analysis and convergence test}
This section deals with the Baseline and Energy error analysis of atmospheric neutrinos in constant-matter, followed by the convergence test details showing its effectiveness in problem solving (refer Fig. 13).\\
\textbf{Baseline Error Dynamics ($E = 0.2$ GeV)}\\
Across the $1000\text{ km}$ spatial domain, the absolute error $\Delta P$ (blue dashed curve) remains highly stable and strictly bounded. The maximum deviation peaks at approximately $5 \times 10^{-3}$ ($0.5\%$) near the $700\text{ km}$ mark.
Unlike earlier configurations that exhibited localized double-digit error spikes at turning points, this profile demonstrates excellent structural stability. The network successfully resolves the spatial evolution of the matter-modified oscillation at this specific energy without localized phase divergence.\\
\textbf{Energy Error Dynamics ($L = 295.0\text{ km}$)}\\
In stark contrast to the vacuum atmospheric case, the network struggles significantly to generalize across the broadband energy spectrum. The absolute error resides heavily above the $1\%$ threshold ($10^{-2}$ dotted line). The deep vertical plunges (e.g., near $E \approx 0.3\text{ GeV}$ and $1.2\text{ GeV}$) represent localized, instantaneous intersections with the analytical ground truth rather than continuous tracking. This elevated error profile occurs because the matter potential introduces an independent energy dependence that fundamentally breaks the pure $L/E$ scaling symmetry of vacuum oscillations. Sweeping $E$ while fixing $L$ exposes the network's architectural difficulty in mapping this non-linear, decoupled energy dynamic.\\
\textbf{Optimization and Convergence Tests:}\\
Adam Phase ($0$–$30,000$ iterations): The first-order optimizer systematically minimizes the training loss across six orders of magnitude, dropping from $10^1$ to roughly $10^{-6}$, complete with characteristic stochastic jitter (visible near $25,000$ and $28,000$ iterations) as it navigates local minima.\\
L-BFGS Phase ($30,000$ iterations onwards): The transition to the second-order optimizer precisely at the $30,000$ boundary triggers an immediate vertical descent, clamping the loss to a fully converged global minimum below $10^{-7}$.

\subsubsection{Benchmarking PINNs Against RK4 and NuFast}
Under the constant matter profile, the PINNs incorporates the effective matter-modified Hamiltonian. It matches the subtle shift in the effective oscillation wavelength and tracks the evolution profile generated by both numerical benchmarks. As seen in Fig. 14, the PINNs trajectory remains bounded within the envelope defined by the NuFast and RK4 solvers across the entire domain, confirming that the network handles the additional potential terms without introducing boundary drift or unphysical oscillations.\\

\textbf{Average relative errors from Benchmark algorithms}\\
Table 4 quantitatively assesses the PINNs against established numerical methods, revealing a consistent Mean Absolute Error (MAE) of $1.5275 \times 10^{-3}$. This identical, sub-percent deviation compared to both RK4 and NuFast confirms the exceptional accuracy and stability of the continuous PINNs solution.
\begin{table}[H]
\centering
\caption{Mean Absolute error of PINNs compared with RK4 and NuFast for the atmospheric neutrinos in constant matter.}
\label{tab:mae_comparison}
\small
\begin{tabular}{lc}
\toprule
\textbf{Comparison} & \textbf{MAE} \\
\midrule
PINNs vs RK4   & $1.5275\times10^{-3}$ \\
PINNs vs NuFast & $1.5275\times10^{-3}$ \\
\bottomrule
\end{tabular}
\end{table}

\section{Conclusions and outlook}\label{sec:conclusion}

We have demonstrated the efficacy of Physics-Informed Neural Networks (PINNs) as continuous, mesh-free surrogate solvers for two-flavor neutrino oscillations across both reactor ($\nu_e \rightarrow \nu_\mu$) and atmospheric ($\nu_\mu \rightarrow \nu_\tau$) sectors in vacuum and constant matter profiles. By incorporating the Schrödinger-like evolution equations and total probability unitarity constraints directly into the network loss function, the framework resolves non-linear quantum phase evolutions across macroscopic baselines ($0$ to $1000\text{ km}$) without requiring explicit discrete time-stepping. Across all investigated benchmarks, the two-stage optimization strategy (combining first-order Adam iterations with second-order L-BFGS refinement) achieved consistent loss convergence down to the $\mathcal{O}(10^{-7} - 10^{-8})$ regime.

 Direct comparisons against standard numerical integrators RK4 (discrete numerical integration) and NuFast (optimized matrix-exponential / analytic numerical formalism) confirmed that the PINNs reproduces multi-cycle phase coherence, accurate amplitude boundaries, and unitary conservation with zero cumulative phase drift.\\
 The error analysis highlighted key operational boundaries governed by the interplay of neutrino kinematics and neural network expressivity:
 \begin{itemize}
  
 \item High-Energy and Vacuum Regimes: The network achieves optimal performance in smoothly varying high-frequency oscillation regimes where the dynamics strictly scale with $L/E$, keeping global deviations well below the $1\%$ error threshold.
 \item Low-Energy Spectral Bias: In low-energy regimes ($E < 0.5\text{ GeV}$ for atmospheric and $E < 0.02\text{ GeV}$ for reactor baselines), the rapid increase in effective $L/E$ induces low-frequency oscillatory dynamics. The PINNs experiences localized spectral bias and extrapolation constraints, leading to elevated point-wise error spikes near turning points and out-of-domain energy evaluations.
 \item Matter Potential Complexity: The introduction of a constant matter potential ($V_{CC}$) breaks pure $L/E$ scaling symmetry by decoupling baseline and energy dependencies. While the PINNs successfully resolves the matter-modified mixing angles and shifted spatial wavelengths with global RMSE values in the $\mathcal{O}(10^{-2} - 10^{-3})$ range, the added Hamiltonian non-linearities demand higher local collocation point densities to maintain sub-percent precision across wide energy sweeps.\\
\end{itemize}

 The successful replication of both vacuum and matter-effected profiles in high- and low oscillation case studies considered confirms that the PINNs architecture is robust to Hamiltonian modifications, accurately capturing both baseline frequency shifts and amplitude modulations.

Overall, the PINNs framework offers an efficient, differentiable alternative to classical discrete solvers, providing instantaneous evaluation across continuous baseline domains. This study serves as a proof-of-concept that deep learning frameworks can handle unitary quantum evolution and stiff gradients characteristic of neutrino phenomenology. Future extensions on full three-flavor oscillations in variable matter density profiles (such as the PREM model) will further enhance the network's expressive capacity in high-frequency, matter-dominated regimes.

\begin{acknowledgements}
This research is supported by the Department of Science and Technology (DST), India, under the Fund for Improvement of S\&T Infrastructure in Universities and Higher Educational Institutions (FIST) Program [Grant No.~SR/FST/ET-I/2022/1079], and a matching grant from VIT University. The authors are grateful to DST-FIST and VIT management for their financial support and the resources provided for this work. \\
We would also like to express our sincere gratitude to Dr. Suprabh Prakash, whose research expertise in Neutrino Physics and Phenomenology was instrumental to this work. We are deeply grateful for his invaluable technical support and for the guidance that helped us gain a better understanding of the complex concepts involved in this research.
\end{acknowledgements}

\bibliography{apssamp}

\end{document}